\documentstyle[preprint,revtex]{aps}
\begin{document}
\draft
\preprint{WIS--93/84/Aug.--PH}

\begin{title}
$\Phi_0$ -- Periodic Aharonov--Bohm Oscillations \\Survive Ensemble Averaging.
\end{title}

\author{Alex Kamenev$(^*)$ and Yuval Gefen$(^*)(^{**})$}

\begin{instit}
$(^*)$Department of Physics, The Weizmann Institute of Science,
Rehovot 76100, Israel.
\newline
$(^{**})$Laboratoire de Physique des Solides, Associe an CNRS, Bat 510,
Universite  Paris--Sud, 91450, Orsay, France.
\end{instit}

\begin{abstract}
We have demonstrated that $\Phi_0$ periodic Aharonov--Bohm
oscillations
measured in a ensemble of rings may survive after ensemble averaging
procedure. The central point is the difference  between the preparation
stage of the ensemble and the subsequent measurement stage. The robustness of
the
effect under finite temperature and non--zero charging energy of rings
is discussed.
\end{abstract}

\pacs{PACS. 05.30, 73.35,  72.10 }

\section{Introduction}
Some of the more interesting results to emerge from the extensive research on
quantum mesoscopic systems has been related to the study of multiply connected
conductors, threaded by an Aharonov--Bohm magnetic flux \cite{Imry86a}.
Physical observables
measured in such systems, are in general, periodically oscillating with the
flux. This applies to transport, as well as to thermodynamic quantities
(including, {\em e.g.}, non--dissipative persistent currents that have been
shown to exist in such systems). Employing gauge invariance, one may show
that these oscillations must be $\Phi_0=hc/e$ periodic. Under most
averaging procedures (most notably impurity averaging), the odd harmonics of
the periodic signal are suppressed, and the effective periodicity turns out
to be \cite{Imry86a,Gefen84}
$\Phi_0/2$.

We have recently \cite{Kamenev93}
pointed out that this period halving depend strongly on the
averaging procedure employed. One is commonly concerned with a set of
macroscopically similar systems ({\em e.g.}, metallic rings all having the
same size, shape and impurity concentration) which are, though,
microscopically distinct ({\em e.g.}, characterized by different microscopic
impurity configurations). In order to systematically compare different types
of ensembles, we had to distinguish between the preparation stage of a given
ensemble and the subsequent measurement stage, where it responds to an
external perturbation. The ensemble we termed Grandcanonical--Canonical
(GC--C) has turned out to be of particular interest. At the preparation stage
each member of the ensemble (``ring'') subject to an external Aharonov--Bohm
flux $\overline{\Phi}$, is brought to equilibrium with an external reservoir
(at temperature $\overline{T}$ and chemical potential $\overline{\mu}$).
That is, as far as preparation is concerned, this ensemble is Grandcanonical.
Subsequently the flux is varied from $\overline{\Phi}$ to $\Phi$ (the
latter may represent either a new static value of the flux or a time
dependent signal), while the number of electrons on each ring is constrained
to remain constant. At the stage of response the ensemble is kept under
canonical conditions. We have been able to express the {\em average} GC--C
persistent current, $I_{\overline{\Phi}}(\Phi)$ in terms of
the average Canonical--Canonical (C--C) current, $I_{cc}(\Phi)$
\cite{Imry90}.
The latter refers to a
situation where not only measurement, but also preparation is done under
canonical conditions. For example, the number of electrons we assign to each
ring is  a randomly selected number in some interval, uncorrelated with the
particular energy spectrum of the ring. The relation we have found is
\cite{Kamenev93}
\begin{equation}
     I_{\overline{\Phi}}(\Phi)=I_{cc}(\Phi)
                         -I_{cc}\left(\frac{\Phi+\overline{\Phi}}{2}\right)
                         -I_{cc}\left(\frac{\Phi-\overline{\Phi}}{2}\right).
                                                             \label{old}
\end{equation}
Thus, although $I_{cc}(\Phi)$ was earlier found to be $\Phi_0/2$ periodic
\cite{Imry90},
the invariance under $\Phi\rightarrow\Phi+\frac{\Phi_0}{2}$ is broken for the
GC--C ensemble, as the preparation flux $\overline{\Phi}$ needs not satisfy
any special symmetry condition. Indeed $I_{\overline{\Phi}}(\Phi)$ is
$\Phi_0$ periodic rather than $\Phi_0/2$ periodic !

It is quite remarkable that all experimental evidence to date supports the
common wisdom that ensemble averaged quantities should be $\Phi_0/2$
periodic. One may raise the question why our $\Phi_0$ periodic prediction has
not been observed so far and whether -- beyond its academic interest --
$\Phi_0$ periodic GC-C current may be connected with any experiment. This
question is of evident interest as the difference between C-C and GC-C
ensembles is qualitative and involves a change in the symmetry of the
measured quantity.

Our earlier analysis of the GC-C ensemble \cite{Kamenev93}
which yielded as a result $\Phi_0$
periodic {\em averaged} signals was confined to rather simplified and
idealized systems: we have considered a zero temperature situation in the
absence of any Coulomb interaction. It is the purpose of the present note to
examine the robustness of this theory (hence of the $\Phi_0$ periodic
signals) against two factors that play a major role in any experimentally
realizable system, namely finite temperature and Coulomb energy. Our two main
results are:

(1) We derive an expression  for the GC-C flux dependent
persistent current both when the preparation temperature, $\overline{T}$, and
the measurement temperature, $T$ (at the stage where the system's response to
an external perturbation is measured) are  \newline finite
\cite{foot1}. We find that
the various harmonics of the $\Phi_0$ periodic signal are washed out as $T$
is increased, the first harmonic being the slowest to die out (Fig.\ (1a)).
The harmonics also depend on the
preparation temperature, $\overline{T}$. Whereas even harmonics will not die
out in the limit of large $\overline{T}$, the odd ones are suppressed as a
power low in $\overline{T}/E_c$ if $\overline{T}\ll E_c$ and become
exponentially small for $\overline{T}\gg E_c$ (Fig's.\ (1b),(1c)).
Here $E_c$ is the Thouless correlation energy of the system:
$E_c = D/L_x^2,$
where $D$ is the system diffusivity and $L_x$ is the ring perimeter.
This behaviour is summarized in Eq.\ (\ref{fin}) and Fig.\ (1).

(2) The charging energy, $E_Q$, of the rings plays a crucial role at the stage
of preparation. We find that the larger $E_Q$ the further we deviate from
GC preparation conditions. The amplitude of the odd harmonics
of the periodic Aharonov--Bohm signal are consequently suppressed by a factor
\begin{equation}
     \frac{\Delta}{\Delta+2 E_Q},
                                                             \label{factor}
\end{equation}
where $\Delta$ is a mean level spacing of the system.
This factor is practically temperature independent.

We thus conclude that under certain experimental requirements and within a
certain temperature range (see the discussion below) $\Phi_0$ periodic
{\em averaged} Aharonov--Bohm oscillations are observable.

\section{Temperature dependence of the GC--C current}
Let us first neglect the charging energy. Extending the
analysis of Ref.\ \cite{Kamenev93}
to finite temperatures, we consider an ensemble of
macroscopically equivalent rings, each having a fixed, flux independent
number of electrons. The current measured at temperature $T$ and flux
$\Phi$ is (cf. Eq.\ (4) of Ref.\ \cite{Kamenev93})
\begin{equation}
I_{\overline{\Phi}}(\Phi)=\langle I^{GC}_T(\Phi)\rangle+
\langle\frac{\partial I^{GC}_T(\Phi)}{\partial\overline{\mu}}
\delta\mu_{T,\overline{T}}(\Phi,\overline{\Phi})\rangle ,
                                                             \label{Iexp}
\end{equation}
where  angular brackets $\langle\ldots\rangle$  stand for the
ensemble averaging.
The first term on the r.h.s. is the vanishingly small averaged
grand canonical current to be neglected hereafter. Let the (sample specific)
particle number be $N_{\overline{T}}(\overline{\Phi})$ (where
$\overline{T}$ and $\overline{\Phi}$ are the preparation temperature and
flux respectively). The variation of the chemical potential under
{\em canonical} measurement conditions is given by
\begin{equation}
\sum_{n} f_T(\epsilon_n(\Phi) - \overline{\mu}
- \delta\mu_{T,\overline{T}}(\Phi,\overline{\Phi})) =
N_{\overline{T}}(\overline{\Phi}),
                                                             \label{cond}
\end{equation}
where $\delta\mu_{\overline{T},\overline{T}}(\overline{\Phi},\overline{\Phi})
\equiv 0$ (by our choice of the preparation procedure),
$T$ and $\Phi$ being the measurement temperature and flux respectively. Here
$f_T$ is the standard Fermi--Dirac distribution function at temperature $T$.
Defining the grand canonical particle number as
\mbox{$N_{T}(\Phi) = \sum_{n} f_T(\epsilon_n(\Phi) - \overline{\mu})$},
we find after expanding Eq.\ (\ref{Iexp}) to  first power in $\delta\mu$
\begin{equation}
\delta\mu_{T,\overline{T}}(\Phi,\overline{\Phi}) =
\Delta\left[N_{\overline{T}}(\overline{\Phi})-N_{T}(\Phi)\right].
                                                             \label{chemp}
\end{equation}
Using a simple thermodynamic relation \cite{Imry90}
$\partial I^{GC}_T(\Phi)/\partial\overline{\mu} =
\partial N_T(\Phi)/\partial\Phi$ and substituting it, together with
Eq.\ (\ref{chemp}), into Eq.\ (\ref{Iexp}), we obtain for the GC-C current.
\begin{equation}
I_{\overline{\Phi}}(\Phi)=
-\frac{1}{2}\Delta\frac{\partial}{\partial\Phi}
\left[\langle N_T^2(\Phi)\rangle-
2\langle N_T(\Phi)N_{\overline{T}}(\overline{\Phi})\rangle \right].
                                                             \label{GCC}
\end{equation}
We next evaluate the correlator
\begin{equation}
K_{T,\overline{T}}(\Phi,\overline{\Phi},\gamma)=
\langle N_T(\Phi)N_{\overline{T}}(\overline{\Phi})\rangle,
                                                             \label{corr}
\end{equation}
where the dependence on the inelastic broadening $\gamma$ has been introduced
explicitly. Within perturbation theory we find
\begin{equation}
K_{T,\overline{T}}(\Phi,\overline{\Phi},\gamma)=
\Re\int d\epsilon K_{0,0}(\Phi,\overline{\Phi},\gamma-i\epsilon)
\xi_{T,\overline{T}}(\epsilon),
                                                             \label{res}
\end{equation}
where the thermal function is \cite{foot2}
\begin{equation}
\xi_{T,\overline{T}}(\epsilon)=
-\frac{\partial^2}{\partial\epsilon^2}
\int dx f_T(x+\frac{\epsilon}{2})f_{\overline{T}}(x-\frac{\epsilon}{2})
                                                             \label{xi}
\end{equation}
and $K_{0,0}(\Phi,\overline{\Phi},\gamma)$ is a well-known zero temperature
correlator, consisting to leading order of a sum of Diffuson and Cooperon
contributions \cite{foot3}.
In terms of the zero temperature C-C persistent current
$I_{cc}(\Phi,\gamma)$, the GC-C current is given by
\begin{eqnarray}
I_{\overline{\Phi}}(\Phi)=
&&\Re\int d\epsilon I_{cc}(\Phi,\gamma-i\epsilon)
\xi_{T,T}(\epsilon)-\nonumber \\
&&\Re\int d\epsilon \left[
I_{cc}\left(\frac{\Phi+\overline{\Phi}}{2},\gamma-i\epsilon\right)+
I_{cc}\left(\frac{\Phi-\overline{\Phi}}{2},\gamma-i\epsilon\right) \right]
\xi_{T,\overline{T}}(\epsilon).
                                                             \label{fin}
\end{eqnarray}

The first term on the r.h.s of Eq.\ (\ref{fin}) is the (measurement
temperature dependent)
C-C current, which contains only even harmonics. Most interesting, it does not
depend on the preparation temperature. The second term on the r.h.s of
Eq.\ (\ref{fin}) gives rise to both odd and even harmonics. These are
suppressed as the preparation temperature is increased (corresponding to the
broadening of $\xi$). As a result the odd harmonics are suppressed with the
preparation temperature, unlike the even ones \cite{foot4}.
The temperature dependence of the four low harmonics is shown in Fig.\ (1).

\section{Effect of charging energy}
Incorporating Coulomb interactions is crucial for a correct description of
the thermodynamics of small rings. We now account for the characteristic
charging energy $E_Q=\frac{e^2}{2C}$, C being the capacitance (see below).

The expression for the sample specific particle number
$N_{\overline{T}}(\overline{\Phi})$ (Eq.\ (\ref{cond})) is now replaced by
\begin{equation}
\overline{N}_{\overline{T}}(\overline{\Phi})=
\frac{\sum_N N exp\{-\frac{1}{\overline{T}}[
F(N,\overline{T},\overline{\Phi})-\overline{\mu}N + E_Q (N-N_0)^2]\}}
{\sum_N  exp\{-\frac{1}{\overline{T}}[
F(N,\overline{T},\overline{\Phi})-\overline{\mu}N + E_Q (N-N_0)^2]\}}.
                                                             \label{Nbar}
\end{equation}
Here $F(N,\overline{T},\overline{\Phi})$ is the free energy of a ring with
$N$ particles (in the absence of any charging energy); $N_0$ represents the
charge (in electron units) of the positive background. At high enough
temperature one may evaluate the sum, Eq.\ (\ref{Nbar}), employing a saddle
point approximation. We obtain
\begin{equation}
\frac{\partial F(N,\overline{T},\overline{\Phi})}{\partial N}
\left|_{N=\overline{N}}\right.
-\overline{\mu} +2 E_Q (\overline{N}-N_0) = 0.
                                                             \label{sp}
\end{equation}
Here
$\frac{\partial F}{\partial N}\left|_{\overline{N}}\right.\equiv
\mu(\overline{N},\overline{\Phi})$, is to be determined from
\begin{equation}
\sum_{n} f_{\overline{T}}(\epsilon_n(\overline{\Phi})
- \mu(\overline{N},\overline{\Phi})) = \overline{N}
                                                             \label{newc}
\end{equation}
($\epsilon_n$, are the single particle energies in the absence of charging
energy).
The particle number $\overline{N}$ is to be determined self-consistently from
Eq.\ (\ref{sp}) and Eq.\ (\ref{newc}). Expanding Eq.\ (\ref{newc}) to the
leading order in $\mu-\overline{\mu}$ and
$\overline{N}-N_{\overline{T}}(\overline{\Phi})$ (cf. Eq.\ (\ref{cond})) and
substituting it in Eq.\ (\ref{sp}) we find
\begin{equation}
\overline{N}_{\overline{T}}(\overline{\Phi})=
\frac{1}{\Delta+ 2 E_Q}
[\Delta N_{\overline{T}}(\overline{\Phi}) + 2 E_Q N_0].
                                                             \label{find}
\end{equation}
Repeating the procedure outlined above (for the $E_Q=0$ scenario), namely
substituting Eq.\ (\ref{find})  in Eq.\ (\ref{chemp}) and that, in turn, in
Eq.\ (\ref{Iexp}) we obtain for the current
\begin{equation}
I_{\overline{\Phi}}(\Phi)=
-\frac{1}{2}\Delta\frac{\partial}{\partial\Phi}
\left[\langle N_T^2(\Phi)\rangle-
\frac{2\Delta}{\Delta+2E_Q}
\langle N_T(\Phi)N_{\overline{T}}(\overline{\Phi})\rangle \right],
                                                             \label{GCCQ}
\end{equation}
which, for $E_Q=0$, reduces to Eq.\ (\ref{GCC}).
Eq.\ (\ref{GCCQ}) is our main result. Two remarks are now due:

(i) The derivation of this equation, following a saddle point approximation
(Eq.\ (\ref{sp})), assumes that many terms in the grand canonical sums
(Eq.\ (\ref{Nbar})) are almost equally important. A necessary condition for
that is that \mbox{$E_Q \ll \overline{T}$} \cite{foot5}.
In the low preparation temperature
limit, \mbox{$E_Q \gg \overline{T}$}, there is generically only one term ($N$)
in
the grand canonical sum which is of importance. It is this $N$ rather than the
solution of the saddle point equation, $\overline{N}$ (cf. Eq.\ (\ref{find})),
which represents the actual number of electrons in the system. Since the
typical preparation flux $\overline{\Phi}$ dependent meander of a single
electron level is \mbox{$\sim\Delta (\ll E_Q)$}, we expect that for
\mbox{$E_Q \gg \overline{T}$} the charging energy indeed quenches the
$\overline{\Phi}$ dependence of the initial particle number, leading to
{\em canonical} preparation condition. Only a fraction $\sim\frac{\Delta}
{E_Q}$ of the ensemble members should still have $\overline{\Phi}$ sensitive
$N_{\overline{T}}(\overline{\Phi})$, qualitatively satisfying the
$\frac{\Delta}{\Delta + 2 E_Q}$ factor in Eq.\ (\ref{GCCQ}) even at low
$\overline{T}$.

(ii) The magnitude of the {\em odd} harmonics in $\Phi$, calculated from the
second term on r.h.s. of Eq.\ (\ref{GCCQ}), is by a factor
$\Delta/(\Delta+2E_Q)$
smaller compared with $E_Q=0$ case. The temperature dependence of the
harmonics remains practically the same as we discussed in the previous
sections.

\subsection{ What is the capacitance}
In order to maximize the odd  harmonic effect discussed above, one should try
to minimize the magnitude of charging energy $E_Q$. We recall that the typical
experiment will consist of coupling the system to a ``particle reservoir'' (a
large conductor), and then cutting it off from the reservoir. For a two
dimensional electron gas this may be achieved  by varying a gate
voltage, as is shown schematically in  Fig.\ (2). The
similar scheme has already been realized experimentally (with a single ring
only) cf. Ref.\ \cite{Mailly93}.
The ring -- reservoir capacitance is quite large (owing to the very small
distance between the ring and the electron gas
at the moment when they are just being decoupled from each other).
We use the parameters given in Ref.\ \cite{Mailly93} to estimate the
reduction factor Eq.\ (\ref{factor}). The circumference of the ring $L=10 \mu
m$, its width $W=0.16 \mu m$, the distance between the ring and the
conductive substrate $d=72 nm$ and the Fermi wavelength $\lambda_F=42 nm$.
With these parameters we estimate the level spacing to be
$\Delta=4.5\times 10^{-6} eV$ and the charging energy as
$E_Q=2.5\times 10^{-6} eV$,
implying a reduction factor of the order of 0.5.

\section{Conclusion}
We have thus demonstrated here that although the observation of
$\Phi_0$ periodic {\em ensemble averaged} signal requires
more care in devising the experimental setup than the observation of a
$\Phi_0/2$ periodic signal, it is
nonetheless possible. The two main requirements are: (i) the preparation
temperature $\overline{T}$, as well as the measurement temperature $T$ should
be smaller or of the order of Thouless energy $E_c$; (ii) the charging energy
of the system $E_Q$ should not exceed much the mean level spacing $\Delta$.
Observation of $\Phi_0$ periodic oscillations is facilitated provided the
ensemble ``remembers''  the
preparation conditions (flux $\overline{\Phi}$), which breaks the symmetry
with respect to \mbox{$\Phi\rightarrow\Phi+\frac{\Phi_0}{2}$}.
We suggest that  differences between GC--C and C--C
conditions (grand canonical versus canonical preparation)
should be observable for other physical uantities defined for
finite size systems as well.

\section{Acknowlegments}
We acknowledge discussions with U. Meirav and D. Mailly on the experimental
realizability of these predictions. Y.G. acknowledges the hospitality of
H. Bouchiat and G. Montambaux at Universite Orsay, France, where this work
was concluded. This research was supported by the German--Israel
Foundation (GIF) and the U.S.--Israel Binational Science Foundation (BSF).

\figure{The first four harmonics of the GC--C current as a functions of the
measurement temperature $T$ ($\overline{T}=0$) (a); and the preparation
temperature
$\overline{T}$ ($T=0$) (b), ($T=.5E_c$) (c). Index of the harmonics is
indicated. The preparation flux $\overline{\Phi}$ is taken to
be zero in all the cases.}

\figure{Schematic plot of the experimental setup. The rings are etched in a 2D
electron gas, (1); metallic contact (particle reservoir), (2) and the gate,
(3).}


\begin{thebibliography}{10}
\bibitem{foot3}
In Ref.\ \cite{Kamenev93} $K_{0,0}(\Phi,\overline{\Phi},\gamma)$
was denoted by
$\tilde{K}(\Phi,\overline{\Phi})=
K(\Phi-\overline{\Phi})+K(\Phi+\overline{\Phi})$ and given by  the diagram
Fig.\ (1a).
It is straightforward to show that for a quasi one--dimensional ring (``short
cylinder'')
$$K(\Phi,\gamma) =
\frac{2}{\pi^2}\sum_{p=1}^{\infty}\frac{1}{p}
e^{-p\sqrt{\gamma/E_c}} \cos 4\pi p\frac{\Phi}{\Phi_0}.$$.
\bibitem{foot4}
Higher preparation temperature implies that the number of electrons
assigned to each ring is less correlated with its specific spectrum; the
preparation is then closer to being canonical.
\bibitem{foot5}
We note that in that case the solution of the saddle point
equation, Eq.\ (\ref{sp}), $\overline{N}$ may be non--integer. This is of no
real consequence since one may always choose for the particle number the
integer closest to $\overline{N}$.
\end{thebibliography}
\end{document}